\documentclass[a4paper,11pt]{article}
\pdfoutput=1
\usepackage{jcappub}

\usepackage[utf8]{inputenc}

\newcommand{\Trh}{T_\text{rh}}
\newcommand{\Tmax}{T_\text{max}}
\newcommand{\arh}{a_\text{rh}}

\newcommand{\gs}{g_\star}
\newcommand{\gss}{g_{\star s}}
\newcommand{\mdm}{m_\text{DM}}

\title{Dark Matter and Leptogenesis\\from Gravitational Production}

\author[a]{Nicolás Bernal}
\author[b]{and Chee Sheng Fong}
\affiliation[a]{Centro de Investigaciones, Universidad Antonio Nariño\\
Carrera 3 este \# 47A-15, Bogotá, Colombia}
\affiliation[b]{Centro de Ciências Naturais e Humanas,
Universidade Federal do ABC\\
09.210-170, Santo André, SP, Brazil}

\emailAdd{nicolas.bernal@uan.edu.co}
\emailAdd{sheng.fong@ufabc.edu.br}

\abstract{
Since the gravitational interaction is universal, any particle that ever existed, if kinematically accessible, has to be produced through her. We explore the possibility that dark matter is generated purely from gravitational scatterings together with heavy Majorana right-handed neutrinos that are long-lived. Their late decay could inject significant entropy into the thermal bath, diluting both the dark matter abundance and the cosmic baryon asymmetry, thereby imposing various constraints on the reheating dynamics.
Additionally to the entropy injection, long-lived right-handed neutrinos could also be responsible for generating the baryon asymmetry through leptogenesis, and hence establish some nontrivial relations between the dark matter and the right-handed properties, and the reheating dynamics.
}

\begin{document}
\begin{flushright}
    PI/UAN-2021-686FT
\end{flushright}

\maketitle

\section{Introduction} 
Dark matter (DM) reaching chemical equilibrium with the standard model (SM) is usually produced in the early Universe via the WIMP paradigm~\cite{Arcadi:2017kky}.
Such interactions could proceed via standard renormalizable portals, i.e., the Higgs portal, the kinetic mixing portal, or the neutrino portal.
Alternatively, if the interaction rates between the dark and visible sectors are not strong enough, DM can originate non-thermally, for example, through the FIMP mechanism~\cite{McDonald:2001vt, Choi:2005vq, Hall:2009bx, Elahi:2014fsa, Bernal:2017kxu}.

However, even if the aforementioned portals connecting the visible and dark sectors are typically assumed, gravity can be the only mediator involved in the production of dark states.%
\footnote{One could also conceive scenarios where gravity and another portal are effective, see e.g., Refs~\cite{Chianese:2020khl, Chianese:2020yjo}.}
In this scenario, massless SM gravitons mediate between the two sectors, giving rise to the production of the dark sector via $s$-channel 2-to-2 annihilation of SM particles~\cite{Garny:2015sjg, Tang:2017hvq, Garny:2017kha, Bernal:2018qlk, Ahmed:2020fhc}.
Being a gravitational process, its contribution is Planck suppressed and can be dominant for high reheating temperatures $\Trh$.

Even if reheating is commonly approximated as an instantaneous event, the generation of the SM thermal bath out of the inflaton energy density is a continuous process. 
In this case, the reheating temperature $\Trh$ corresponds to the onset of the radiation-dominated era, while the thermal bath can reach a maximal temperature $\Tmax$, in general much higher than $\Trh$~\cite{Giudice:2000ex}.
During this heating epoch, i.e., while $\Tmax \geq T \geq \Trh$, inflatons dominate the energy density of the Universe, and can populate the dark sector via 2-to-2 scatterings mediated again by the $s$-channel exchange of SM gravitons~\cite{Ema:2015dka, Ema:2016hlw, Ema:2018ucl, Mambrini:2021zpp, Barman:2021ugy}.%
\footnote{Recently, another purely gravitational production of DM has received growing attention, corresponding to the Hawking radiation by primordial black holes during its evaporation~\cite{Green:1999yh, Khlopov:2004tn, Dai:2009hx, Fujita:2014hha, Allahverdi:2017sks, Lennon:2017tqq, Morrison:2018xla, Hooper:2019gtx, Chaudhuri:2020wjo, Masina:2020xhk, Baldes:2020nuv, Gondolo:2020uqv, Bernal:2020kse, Bernal:2020ili, Bernal:2020bjf, Auffinger:2020afu, Bernal:2021akf}.
Additionally, the gravitational production can be enhanced in scenarios with extra dimensions, see e.g., Refs.~\cite{Lee:2013bua, Lee:2014caa, Han:2015cty, Rueter:2017nbk, Kim:2017mtc, Rizzo:2018ntg, Carrillo-Monteverde:2018phy, Kim:2018xsp, Rizzo:2018joy, Goudelis:2018xqi, Brax:2019koq, Folgado:2019sgz, Kumar:2019iqs, Folgado:2019gie, Kang:2020huh, Chivukula:2020hvi, Kang:2020yul, Kang:2020afi, Bernal:2020fvw, Bernal:2020yqg}.}

Besides the DM, the baryon asymmetry of the Universe (BAU) constitutes another major puzzle of the SM.
This asymmetry can be dynamically generated via leptogenesis~\cite{Fukugita:1986hr, Fong:2013wr}.
In that case, the decay of Majorana right-handed neutrinos (RHNs) into a Higgs boson and a lepton doublet produces a baryon-minus-lepton ($B-L$) asymmetry that is converted to a baryonic asymmetry by electroweak sphalerons. 
Since the gravitational interaction is universal, besides DM, heavy metastable particles, including the RHNs, can also be produced gravitationally. In this case, if they are long-lived, they can potentially dominate the energy density of the Universe and upon decay, inject entropy and significantly dilute the DM abundance as well as the BAU~\cite{Allahverdi:2020bys}. The only limit is that they should decay before the Big Bang Nucleosynthesis (BBN), at $T = T_\text{BBN} \simeq 4$~MeV~\cite{Sarkar:1995dd, Kawasaki:2000en, Hannestad:2004px, DeBernardis:2008zz, deSalas:2015glj}. In other words, the Universe could be matter-dominated up till the BBN. The effect on DM is straightforward, the DM mass has to be heavier or have a more efficient production mechanism to compensate for the dilution in their abundance. For the BAU, since there is no room to change the nucleon mass, there is a limit on how much dilution is allowed depending on the production mechanism of BAU. In this work, we identify the RHNs as natural candidates of long-lived heavy particles which are produced gravitationally and are also responsible for generating the BAU through leptogenesis.%
\footnote{We have ignored the possibility of producing the RHNs from inflaton decay which depends on a non-gravitational coupling. However, it could also lead to nonthermal leptogenesis, as explored in Refs.~\cite{Lazarides:1991wu, Murayama:1992ua, Kolb:1996jt, Giudice:1999fb, Asaka:1999yd, Asaka:1999jb, Hamaguchi:2001gw, Jeannerot:2001qu, Fujii:2002jw, Giudice:2003jh, Pascoli:2003rq, Asaka:2002zu, Panotopoulos:2006wj, HahnWoernle:2008pq}, with possible connections to DM~\cite{Borah:2020wyc, Samanta:2020gdw, Barman:2021tgt}. 
In other words, we are considering the most conservative case.} 

Considering three families of RHNs, if the reheating temperature is higher than their mass scale, one of them, which we denote $N_f$ (for \emph{feebly} coupled) is allowed to be long-lived while the other two $N_{g\neq f}$ have to be in thermal equilibrium due to the observed neutrino mass scale~\cite{Bernal:2017zvx}. Even if $N_f$ is feebly coupled to the SM sector, they are inevitably produced in the early Universe via inverse decays. In this work we will focus on scenarios where the decay width is suppressed, making this production mechanism subdominant compared to the gravitational production.
We will consider two scenarios: $(i)$ $N_g$ is the main source of leptogenesis while $N_f$ only provides dilution and $(ii)$ $N_f$ dilutes both the DM and the BAU, while being simultaneously the main source of leptogenesis.
In scenario $(i)$, in order not to over-dilute the BAU, one obtains an upper bound on reheating temperature as a function of lifetime of $N_f$ and this in turn limits the dilution of the DM abundance.
In scenario $(ii)$, since the abundance of $N_f$ responsible for leptogenesis is tied to gravitational production just like the DM, there is tight correlation between the DM and $N_f$ mass.

The paper is organized as follows.
In section~\ref{sec:GravProd}, we review in detail the gravitational production of the dark sector (or any heavy particle): i.e., via 2-to-2 annihilations of inflatons and SM particles mediated by the $s$-channel exchange of gravitons.
Section~\ref{sec:dil} is devoted to the potential dilution of the DM and the $B-L$ asymmetry due to the decay of a long-lived RHN.
Such asymmetry is created via leptogenesis and described in section~\ref{sec:lepto}.
The strong constraints imposed by requiring gravitational processes to generate simultaneously the DM and the BAU are presented in section~\ref{sec:DMBAU}.
Finally, in section~\ref{sec:con} our conclusions are presented.

\section{Gravitational Production of the Dark Sector} \label{sec:GravProd}
The number density $n$ of a dark sector state (i.e., DM or RHNs in our case) is given by the Boltzmann equation
\begin{equation} \label{eq:BE0}
    \frac{dn}{dt} + 3H\,n = \gamma\,,
\end{equation}
where $H$  corresponds to the Hubble expansion rate, and $\gamma$ to the production rate density of the given dark sector state.%
\footnote{Self-interactions within the dark sector could also alter the number density, but in this case where they would have a gravitational nature they are subdominant because of a suppression by higher orders of $M_P$~\cite{Bernal:2020gzm}.}
Here we consider the production via 2-to-2 annihilations of inflatons or SM states, via the $s$-channel exchange of SM gravitons.
These two channels will be described in the following subsections.

\subsection{Inflaton Scatterings}
While reheating is commonly approximated as an instantaneous event, the generation of the SM thermal bath out of the inflaton energy density is a continuous process.%
\footnote{We note that SM particles do not necessarily thermalize
instantaneously, and thus the decay products are initially distributed with smaller occupation numbers and harder momenta~\cite{Harigaya:2013vwa, Ellis:2015jpg, Garcia:2018wtq}. However, here we will assume that the SM thermalizes rapidly.}
During reheating, the thermal bath reaches a maximal temperature $\Tmax$ that could be much higher that the reheating temperature $\Trh$, corresponding to the onset of the radiation-dominated era~\cite{Giudice:2000ex}.
It is typically assumed that in this era inflatons behave as nonrelativistic matter and constantly decay into SM radiation.
Therefore, SM radiation is not free and its energy density $\rho_R$ scales as $\rho_R(a) \propto a^{-\frac32}$, with $a$ being the scale factor.
Hence, its temperature scales as
\begin{equation}
    T(a) \propto a^{-\frac38}.
\end{equation}
This implies that the inflaton energy density $\rho_\phi$ is given by
\begin{equation}
    \rho_\phi(T) = \frac{\pi^2\,\gs}{30}\frac{T^8}{\Trh^4}\,,
\end{equation}
where $\rho_\phi(\Trh) = \rho_R(\Trh)$ was assumed, and with $\gs$ being the number of relativistic degrees of freedom contributing to the SM energy density~\cite{Drees:2015exa}.
During reheating, the Hubble expansion rate is dominated by the inflaton, and therefore
\begin{equation}
    H(T) = \frac{\pi}{3} \sqrt{\frac{\gs}{10}}\, \frac{T^4}{\Trh^2\,M_P}\,,
\end{equation}
with $M_P \simeq 2.4 \times 10^{18}$~GeV the reduced Planck mass.

To track the evolution of the dark sector states during the reheating era in which the SM entropy is not conserved due to the decay of the inflaton, Eq.~\eqref{eq:BE0} can be rewritten in terms of the comoving number density $N \equiv n \times a^3$ as
\begin{equation} \label{eq:BE1}
    \frac{dN}{dT} = -\frac{8}{\pi} \sqrt{\frac{10}{\gs}}\, \frac{\Trh^{10}\,M_P}{T^{13}}\, \arh^3\, \gamma\,,
\end{equation}
with $\arh \equiv a(\Trh)$.

The interaction rate density for 2-to-2 production out of nonrelativistic inflatons is given by~\cite{Mambrini:2021zpp}
\begin{equation} \label{eq:scalarfermion}
    \gamma = \frac{\rho_\phi^2}{1024\pi\,M_P^4}\, f\left(\frac{m}{m_\phi}\right) = \frac{\pi^3\,\gs^2}{921600} \frac{T^{16}}{M_P^4\,\Trh^8}\, f\left(\frac{m}{m_\phi}\right),
\end{equation}
where $m$ and $m_\phi$ are the masses of the produced particle and the inflaton, respectively, and where we have defined
\begin{equation}
f(x) \equiv
    \begin{cases}
        \left(1+\frac{1}{2}x^{2}\right)^{2}\sqrt{1-x^{2}} & \mbox{for real scalars},\\[8pt]
        \frac{1}{8}\, x^{2}\left(1-x^{2}\right)^{3/2} & \mbox{for Majorana fermions}.
    \end{cases}
\end{equation}

The yield $Y(T) \equiv n(T)/s(T)$ is defined as a function of the SM entropy density $s \equiv \frac{2\pi^2}{45}\gss\, T^3$, with $\gss(T)$ being the number of relativistic degrees of freedom contributing to the SM entropy~\cite{Drees:2015exa}, and can be computed by integrating Eq.~\eqref{eq:BE1} together with Eq.~\eqref{eq:scalarfermion} in the range $\Tmax \geq T \geq \Trh$.
The yield $Y_0$ at late times ($T \ll \Trh$) is therefore
\begin{equation} \label{eq:Y0}
    Y_0 = Y(\Trh) = \frac{N(\Trh)}{s(\Trh)\, \arh^3}
    \simeq \frac{\gs^2}{20480\gss} \sqrt{\frac{10}{\gs}}\, \frac{\Tmax^4}{M_P^3\,\Trh}\, f\left(\frac{m}{m_\phi}\right).
\end{equation}

\begin{figure}
	\centering
	\includegraphics[scale=0.51]{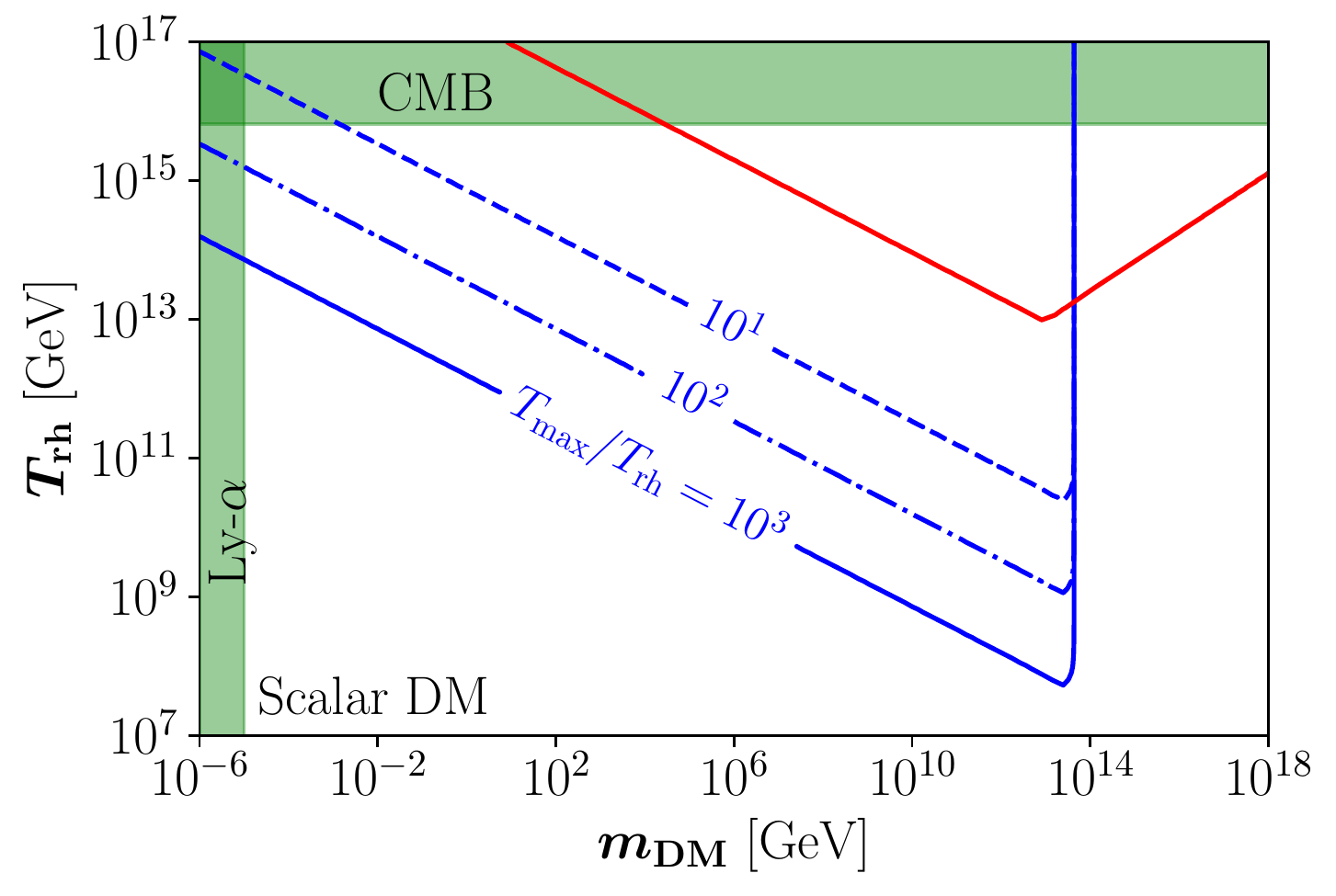}
	\includegraphics[scale=0.51]{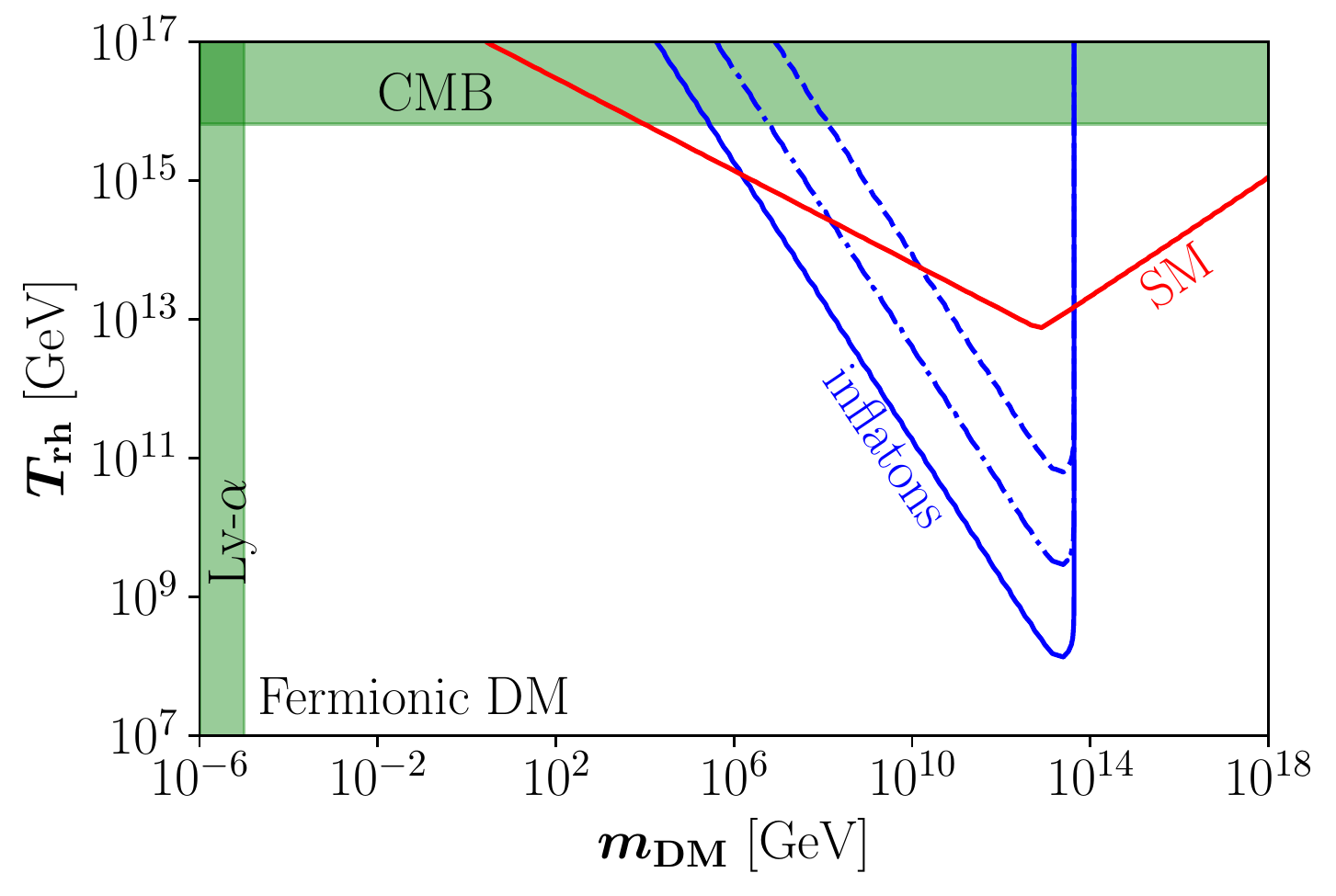}
    \caption{Contour lines for the parameter space reproducing the observed DM relic abundance, for scalar (left panel) and fermionic (right panel) DM. 
    Blue lines correspond to the DM production via gravitational annihilation of inflatons, for $\Tmax/\Trh=10^3$ (solid), $10^2$ (dash-dotted) and $10^1$ (dashed), assuming $m_\phi = 3 \times 10^{13}$~GeV.
    Red lines correspond to the gravitational freeze-in due to annihilation of SM particles.
    Green areas are in tension with CMB and Lyman-$\alpha$ observations.
    }
	\label{fig:DMprod}
\end{figure}

The scattering of inflatons could produce DM particles.
In that case, to match the observed DM abundance $\Omega h^2 \simeq 0.12$, the DM yield has to be fixed so that $\mdm Y_0 = \Omega h^2 \frac{1}{s_0} \frac{\rho_c}{h^2} \simeq 4.3 \times 10^{-10}$~GeV, where $\rho_c \simeq 1.1 \times 10^{-5} h^2$~GeV/cm$^3$ is the critical energy density, and $s_0 \simeq 2.9 \times 10^3$~cm$^{-3}$ is the entropy density at present~\cite{Aghanim:2018eyx}.
Figure~\ref{fig:DMprod} shows with blue lines the parameter space that reproduces the observed DM abundance, for scalar (left panel) and fermionic (right panel) DM generated by the gravitational annihilation of inflatons.
We have taken the ratio $\Tmax/\Trh=10^3$ (solid), $10^2$ (dash-dotted) and $10^1$ (dashed), and assumed $m_\phi = 3\times 10^{13}$~GeV.%
\footnote{If one does not fix an inflationary and heating model, $m_\phi$, $\Tmax$ and $\Trh$ are free parameters.
Here we will assume $m_\phi = 3 \times 10^{13}$~GeV and often $\Tmax/\Trh = 10^3$, inspired by the successful Starobinsky inflationary scenario~\cite{Starobinsky:1980te, Bernal:2020qyu}.}
Regions above the lines generate a DM overdensity, overclosing the Universe.
The green upper bands (labeled CMB), corresponding to $\Trh \geq 6.5 \times 10^{15}$~GeV, are in tension with the upper limit of the inflationary scale $H_I \leq 2.5 \times 10^{-5} M_P$~\cite{Akrami:2018odb}.
Finally, the green bands on the left (labeled Ly-$\alpha$) correspond to DM masses lighter than $\sim 10$~keV, in tension with the Lyman-$\alpha$ bound~\cite{Ballesteros:2020adh, DEramo:2020gpr}.

\subsection{SM Scatterings}
Alternatively, DM and RHNs can also be gravitationally produced via the UV freeze-in mechanism, by 2-to-2 annihilation of SM particles mediated by the $s$-channel exchange of a massless graviton.
The corresponding interaction rate density takes the form~\cite{Garny:2015sjg, Tang:2017hvq, Garny:2017kha, Bernal:2018qlk}
\begin{equation}
    \gamma(T) = \alpha\, \frac{T^8}{M_P^4}\,,
\end{equation}
with $\alpha \simeq 1.9\times 10^{-4}$ or $\alpha \simeq 5.5\times 10^{-4}$ for real scalar or Majorana fermions, respectively.
Using the instantaneous decay approximation for the inflaton, Eq.~\eqref{eq:BE0} can be rewritten as
\begin{equation}
    \frac{dY}{dT} = - \frac{\gamma(T)}{H(T)\, T\, s(T)}\,,
\end{equation}
where $H(T) = \frac{\pi}{3} \sqrt{\frac{\gs}{10}} \frac{T^2}{M_P}$ for a radiation-dominated Universe, and admits the analytical solution $Y_0$ for $T \ll \Trh$
\begin{equation} \label{eq:FIlight}
    Y_0 = \frac{45\,\alpha}{2\pi^3\,\gss} \sqrt{\frac{10}{\gs}} \left(\frac{\Trh}{M_P}\right)^3,
\end{equation}
in the case $m \ll \Trh$.
We note that away from the instantaneous decay approximation of the inflaton, the yield is only boosted by a small factor of order $\mathcal{O}(1)$, for an inflaton behaving as nonrelativistic matter~\cite{Garcia:2017tuj, Bernal:2019mhf}.
Instead, if the produced particle is heavier than the reheating temperature (but still lighter than $\Tmax$), it cannot be generated after but during reheating.
In that case, the yield can be computed by integrating Eq.~\eqref{eq:BE1} in the range $\Tmax \geq T \geq m$, and corresponds to
\begin{equation} \label{eq:FIheavy}
    Y_0 = \frac{45\,\alpha}{2\pi^3\,\gss} \sqrt{\frac{10}{\gs}} \frac{\Trh^7}{M_P^3\, m^4}\,.
\end{equation}
Figure~\ref{fig:DMprod} also shows with red lines the parameter space reproducing the observed DM abundance via the UV freeze-in by annihilation of SM particles mediated by gravitons, for scalar (left panel) and fermionic (right panel) DM.
Again, regions above the lines generate a DM overdensity, overclosing the Universe.
We emphasize that this channel is largely independent on $\Tmax$.

\section{Dilution from Long-lived Particles} \label{sec:dil}

\subsection{Generality}

Since gravitational interaction is universal, all types of particles that
ever exist can be produced through it. Depending on the reheating
dynamics, heavy particles can be produced in significant amounts.
Furthermore, if these heavy particles are long-lived, the Universe can
go through a matter-dominated period and their subsequent decays will inject significant entropy and hence dilute the DM as well as the BAU that are produced earlier. Since the gravitational production from thermal bath particles is subdominant compared to through inflatons, we will focus on the latter case.

Let us assume that a long-lived particle $X$ of mass $m_{X}$ is produced
gravitationally through inflatons up till $\Trh$, their final
abundance is given by Eq.~\eqref{eq:Y0}.
In the absence of entropy injection, $Y_{X}\left(T\right)$ is constant
for $T<\Trh$.
Its decay temperature $T_{d}<m_{X}$ corresponds to the SM temperature at which the Hubble expansion rate equals the decay width
\begin{equation} \label{eq:decay_condition}
    \Gamma_{X} = H\left(T_{d}\right),
\end{equation}
where $\Gamma_{X}$ corresponds to the total decay width of $X$, and the Hubble rate is $H\left(T\right)=\sqrt{\frac{\rho\left(T\right)}{3M_{P}^{2}}}$, with the total cosmic energy density given by
\begin{equation}
    \rho\left(T\right) = \frac{\pi^{2}}{30} \gs T^4 + \rho_{X}(T),
\end{equation}
where the first term is the contribution from thermal bath radiation
and the second is the $X$ energy density.
In the case where it is nonrelativistic, its energy density takes the form
\begin{equation}
    \rho_{X}\left(T\right) = m_{X}\, s\left(T\right)\, Y_{X}(T).
\end{equation}
From Eq.~\eqref{eq:decay_condition}, one can solve for $T_{d}$ as
a function of $\Gamma_{X}$, $m_{X}$, $\Trh$, $\Tmax$ and $m_{\phi}$. 
In the following, we will trade $\Gamma_{X}$ for $T_{d}$, and treat $T_{d}$ as a free parameter. In order not to affect BBN, we will require $T_{d} > T_\text{BBN}$.

Next, we will assume that $X$ {\it only} decays to SM particles, independently on which final states.
Hence $X$ is also unavoidably produced from inverse decays. Since $X$ is assumed to be long-lived, it never achieves thermal equilibrium, and its freeze-in abundance from inverse decay is given by~\cite{Agashe:2018cuf}
\begin{equation} \label{eq:YX_fi}
    Y_X^{\rm id} = \frac{405}{8 \pi^4} \frac{g_X}{\gss} \sqrt{\frac{10}{\gs}} \frac{M_P\, \Gamma_X}{m_X^2}\,,
\end{equation}
where $g_X$ is the degrees of freedom of $X$ and we have assumed that the production is completed during the radiation dominated period, i.e., before $X$ starts to dominate the energy density of the Universe.

Later, the decay of $X$ to the SM particles can inject significant entropy to the thermal bath. In the sudden decay approximation, the conservation of energy density implies
\begin{equation}
    \frac{\pi^{2}}{30}\gs T_{d}^{4}+\rho_{X}\left(T_{d}\right) =\frac{\pi^{2}}{30}\gs \widetilde{T}^{4},
\end{equation}
where $\widetilde{T}$ is the temperature of thermal bath radiation
right after $X$ decays. The dilution factor $d$ due to entropy injection
from $X$ will be
\begin{equation} \label{eq:dilution}
    d \equiv \frac{s\big(\widetilde{T}\big)}{s\left(T_{d}\right)} = \left(\frac{\widetilde{T}}{T_{d}}\right)^{3} = \left[1 + \frac{1}{1536}\sqrt{\frac{\gs}{10}}\, \frac{m_X}{T_d} \left(\frac{\Trh}{M_{P}}\right)^{3} \left(\frac{\Tmax}{\Trh}\right)^4 f\left(\frac{m_{X}}{m_{\phi}}\right)\right]^{3/4}.
\end{equation}
For $d\gg1$, the dilution scales as
\begin{equation} \label{eq:dil}
    d \simeq 1.4\times10^{3} \left(\frac{\gs }{106.75}\right)^{3/8} \left(\frac{m_X/T_d}{100}\right)^{3/4} \left(\frac{\Trh}{10^{16}\,{\rm GeV}}\right)^{9/4} \left(\frac{\Tmax/\Trh}{1000}\right)^{3} f^\frac34\left(\frac{m_{X}}{m_{\phi}}\right).
\end{equation}
For a scalar $X$, $f\left(\frac{m_{X}}{m_{\phi}}\right)\sim1$ while
for fermion $X$, we will have an additional suppression factor $f\left(\frac{m_{X}}{m_{\phi}}\right)\sim\frac18\left(\frac{m_{X}}{m_{\phi}}\right)^{2}$.

Using again the sudden decay approximation for $X$, it follows that $\Gamma_X = H(T_d) = H(\widetilde T)$, and hence Eq.~\eqref{eq:YX_fi} can be rewritten as
\begin{equation} \label{eq:YX_fi2}
    Y_X^{\rm id} = \frac{675\, g_X}{4\pi^3\, \gs\, \gss} \bigg(\frac{\widetilde T}{m_X}\bigg)^2.
\end{equation}
It is interesting to note that the production via inverse decay is naturally subdominant with respect to the gravitational production if the $X$ particle is long-lived enough.
Comparing Eqs.~\eqref{eq:Y0} with~\eqref{eq:YX_fi2}, one gets that if
\begin{equation}
    T_d \ll \frac{\pi^2}{480} \left(\frac{\gs^4}{45\, g_X^2} \sqrt{\frac{\gs}{10}}\right)^\frac13 \frac{\Trh}{M_P} \left(\frac{\Tmax}{\Trh}\right)^\frac43 \left(\frac{m_X}{m_\phi}\right)^\frac43 m_X,
\end{equation}
or equivalently if
\begin{equation}
        \frac{T_d}{m_X} \ll 4\times 10^{-3}\, \frac{\Trh}{10^{15}~\text{GeV}} \left(\frac{\Tmax/\Trh}{10^3}\right)^\frac43 \left(\frac{3\times 10^{13}~\text{GeV}}{m_\phi}\right)^\frac43 \left(\frac{m_X}{10^{11}~\text{GeV}}\right)^\frac43,
\end{equation}
the gravitational production is indeed dominant, for the case of fermionic $X$ particles.

\subsection[Long-lived Right-handed Neutrinos $N_f$]{\boldmath Long-lived Right-handed Neutrinos $N_f$}

Now we will identify a candidate of long-lived particle $X$ with
the well-motivated RHNs which are responsible
for generating neutrino masses through type-I seesaw mechanism~\cite{Minkowski:1977sc, Yanagida:1979as, GellMann:1980vs, Mohapatra:1980yp} and
generating the BAU through leptogenesis. Working in the mass basis of RHNs $N_{i}$ ($i$ is the family index) and where the charged
lepton Yukawa coupling is diagonal, the relevant Lagrangian terms
for the model are given by
\begin{equation}
    -{\cal L} \supset \frac{1}{2} m_{N_{i}}\, \overline{N_{i}^{c}}\, N_{i} + \lambda_{\alpha i}\, \overline{\ell_{\alpha}}\, \epsilon\, \Phi^{*} N_{i} + {\rm H.c.},
\end{equation}
where $\ell_{\alpha}$ $(\alpha=e$, $\mu$, $\tau)$ and $\Phi$ are respectively
the SM lepton and Higgs doublets, and $\epsilon$ is the totally antisymmetric
tensor of $SU(2)_{L}$. From the observed light neutrino mass differences,
if the reheating temperature is higher than the mass of RHNs, at least
some of them have to be in thermal equilibrium due to the interactions
mediated by neutrino Yukawa coupling $\lambda_{\alpha i}$~\cite{Bernal:2017zvx}.
If they
are thermalized by the neutrino Yukawa interactions, they will acquire
thermal abundance making the gravitational production irrelevant.
Furthermore, once they are nonrelativistic, they will quickly decay
before dominating the cosmic energy density, resulting in insignificant
entropy production $d \simeq 1$. 

We will consider the scenario motivated by $SO(10)$ grand unified
theories which predicts three RHNs $(i=1$, 2, 3). In this case, two
of them have to be in thermal equilibrium while the other can be
out of thermal equilibrium with the SM~\cite{Bernal:2017zvx}.
After the electroweak symmetry breaking, the light neutrino mass matrix
is given by~\cite{Minkowski:1977sc,Yanagida:1979as,GellMann:1980vs,Mohapatra:1980yp} 
\begin{equation}
    m_{\nu} = -v^{2}\, \lambda\, \hat{M}^{-1}\, \lambda^{T},
\end{equation}
where $v \equiv \left\langle \Phi \right\rangle \simeq 174$~GeV is the vacuum
expectation value of $\Phi$ and $\hat{M} \equiv {\rm diag}(m_{N_{1}}$, $m_{N_{2}}$, $m_{N_{3}})$,
assuming $\left|\lambda_{\alpha i}\right| v \ll m_{N_{i}}$. From the
above, $m_{\nu}$ can be diagonalized by a unitary matrix, identified
with the Pontecorvo-Maki-Nakagawa-Sakata matrix, as $U^{T}\, m_{\nu}\, U = {\rm diag}(m_{1}$, $m_{2}$, $m_{3})$.
From the global fit,
one has $\Delta m_{{\rm sol}}^{2} \equiv m_{2}^{2} - m_{1}^{2} \simeq 7.4 \times 10^{-5}$~eV$^{2}$
and $\Delta m_{{\rm atm}}^{2} \equiv \left|m_{3}^{2} - m_{l}^{2}\right| \simeq 2.5 \times 10^{-3}$~eV$^{2}$~\cite{Esteban:2020cvm}.
Experimentally, two possibilities are allowed: for Normal mass Ordering
(NO), one chooses $m_{3}^{2} - m_{l}^{2} = m_{3}^{2} - m_{1}^{2} > 0$ while
for the Inverse mass Ordering (IO), one takes $m_{3}^{2}-m_{l}^{2}=m_{3}^{2}-m_{2}^{2}<0$. 
As an estimation, for $m_\nu \sim 0.1$ eV, the perturbativity bound $|\lambda| \lesssim 1$ implies 
\begin{equation} \label{eq:pert_bound_M}
    m_{N_i} \lesssim 3\times 10^{14}~{\rm GeV}.
\end{equation}
In the following figures, this will be denoted as ``perturbativity'' bound on the RHN mass scale.

Next, let us consider the possibility that one of the RHNs, denote $N_{f}$, is not thermalized by the neutrino Yukawa interactions. The subscript $f$ on $N_f$ denotes the RHN which is \emph{feebly} coupled and responsible for possible dilution due to its late decay.
We note that it can be a RHN of any generation.
In this case, its abundance will solely be determined by the gravitational
production from inflatons given by the fermionic case of Eq.~\eqref{eq:Y0} with $m = m_{N_f}$. Once produced, it
can have a long lifetime with a decay width given by
\begin{equation}
    \Gamma_{N_f}  =  \frac{\left(\lambda^{\dagger} \lambda\right)_{ff}\, m_{N_{f}}}{8\pi} \equiv \frac{\tilde{m}_{f}\, m_{N_{f}}^{2}}{8\pi\, v^{2}},
\end{equation}
where we have defined the effective neutrino mass as
\begin{equation}
    \tilde{m}_{f} \equiv \frac{\left(\lambda^{\dagger} \lambda\right)_{ff} v^{2}}{m_{N_{f}}}.
\end{equation}
The decay width $\Gamma_{N_{f}}$ is minimized when $\tilde{m}_{f} = m_{1}$
for NO, and $\tilde{m}_{1} = m_{3}$ for IO.
Since there is no lower bound on $m_{1}$ ($m_{3}$) for NO (IO),
we see that $\Gamma_{N_{f}}$ and hence the decay temperature $T_{d}$
can be as small as desired down to the BBN scale.\footnote{For instance, taking $m_{N_f} \sim 10^{8}$~GeV,  Yukawa coupling has to be larger than $\sim 10^{-15}$.  We note that small Yukawa couplings are stable since radiative corrections are proportional to the couplings themselves (i.e., technically natural).}
If $N_{f}$ is responsible for leptogenesis, one will further require
$T_d \gtrsim 130$~GeV such that significant BAU can be induced
before the electroweak sphaleron interactions freeze out~\cite{DOnofrio:2014rug}. 
We emphasize that as small values for $T_d$ are considered, Eq.~\eqref{eq:YX_fi2} is automatically satisfied, and therefore $N_f$ is mainly produced by the scattering of inflatons and not by its inverse decay.
The dilution from its decay is given by the fermionic case of Eq.~\eqref{eq:dilution} with $m_{X} = m_{N_{f}}$.

\subsection{Dilution of Dark Matter from Right-handed Neutrinos}
\begin{figure}
	\centering
	\includegraphics[scale=0.51]{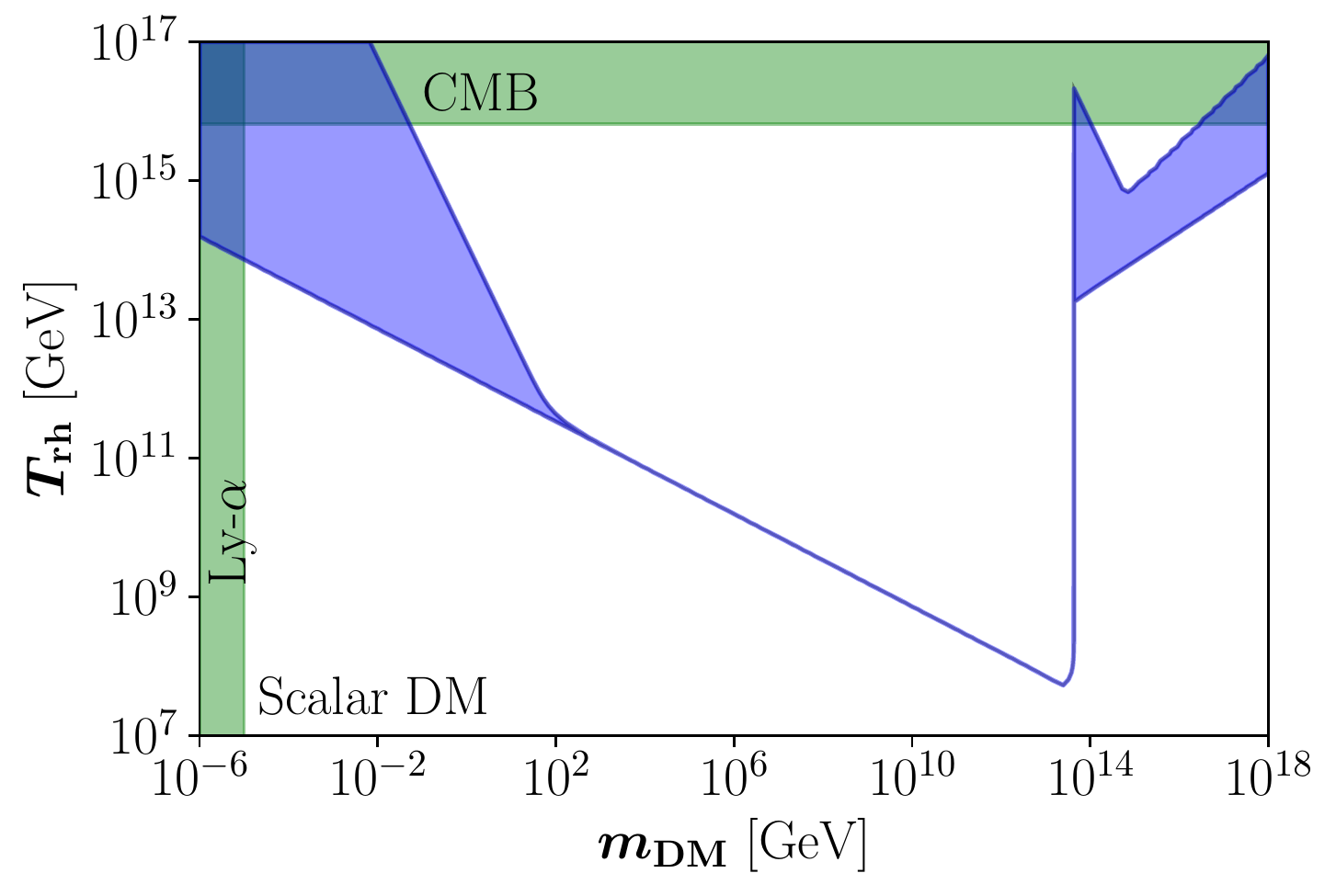}
	\includegraphics[scale=0.51]{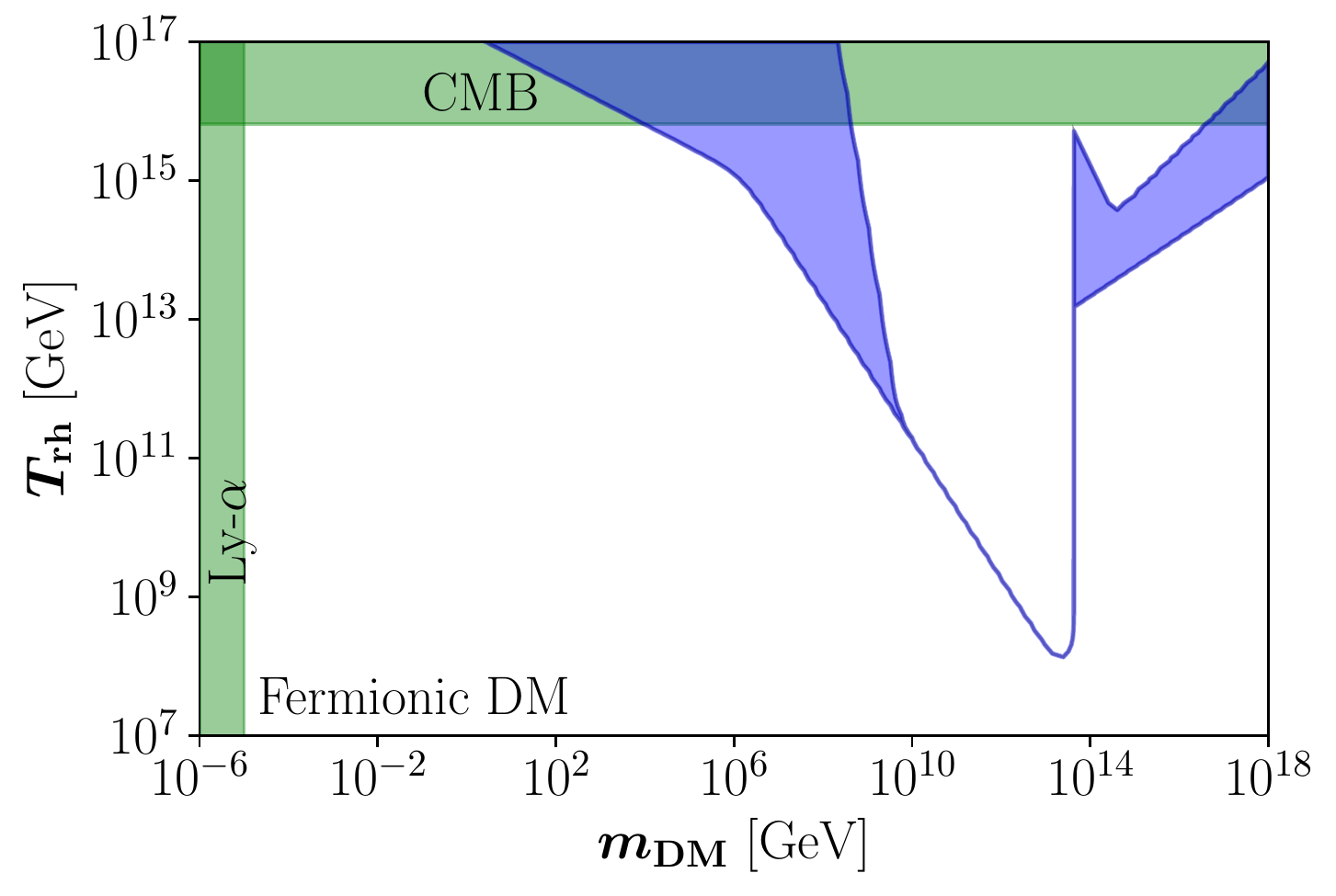}
    \caption{Contour lines for the parameter space reproducing the observed DM relic abundance, for scalar (left panel) and fermionic (right panel) DM, for $\Tmax/\Trh=10^3$, $m_\phi = 3 \times 10^{13}$~GeV, and $m_{N_f} = 10^{12}$~GeV.
    The width of the blue bands show the effect of dilution on the DM abundance.
    Green areas are in tension with CMB and Lyman-$\alpha$ observations.
    }
	\label{fig:DMprod-dil}
\end{figure} 
Figure~\ref{fig:DMprod-dil} shows the parameter space that reproduces the observed DM abundance, for scalar (left panel) and fermionic (right panel) DM generated by gravitational annihilation of inflatons and SM states.
We have assumed $\Tmax/\Trh = 10^3$, $m_{N_f} = 10^{12}$~GeV, and $m_\phi = 3 \times 10^{13}$~GeV.
The thickness of the blue bands corresponds to the effect of dilution due to the decay of a RHN.
The lower border of the band comes from the case without dilution, whereas the upper one from the maximal dilution, when $T_d = T_\text{BBN}$.
As expected from Eq.~\eqref{eq:dil}, this effect (and therefore the width of the bands) is maximized for high values of $\Trh$.
For $\Trh \lesssim 10^{11}$~GeV the dilution parameter $d \simeq 1$, and therefore the bands collapse to lines.
We note that the change of slope at $\mdm \sim 10^{15}$~GeV reflects the two regimes $\mdm < \Trh$ and $\mdm > \Trh$, in Eqs.~\eqref{eq:FIlight} and~\eqref{eq:FIheavy}.
As an example, the final DM yield produced by the annihilation of inflatons once considered the dilution by entropy injection, is
\begin{equation} \label{eq:DM}
    Y_0 = \frac{Y(\Trh)}{d}
    \simeq \frac{\gs}{\gss} \left[\frac{27}{2048} \sqrt{\frac{\gs}{10}} \frac{\Tmax^4\,T_d^3\, m_\phi^6}{M_P^3\, m_{N_f}^9\, \Trh}\right]^\frac14 \times
    \begin{cases}
        1 &\quad \text{for scalar DM,}\\[8pt]
        \frac18 \left(\frac{\mdm}{m_\phi}\right)^2 &\quad \text{for fermionic DM,}
    \end{cases}
\end{equation}
in the case where $d \gg 1$.

\section{\boldmath $B-L$ Asymmetry} \label{sec:lepto}

We will consider two possible leptogenesis scenarios to generate a $B-L$ asymmetry.
In scenario $(i)$, $N_f$ is not responsible for the $B-L$ asymmetry generation but only for the dilution of the asymmetry generated at higher scales by $N_{g \neq f}$, which decay promptly as they become nonrelativistic. In scenario $(ii)$, $N_f$ simultaneously generates the $B-L$ asymmetry and dilutes by injecting entropy.
Here only the previously mentioned limiting scenarios will be considered. Though intermediate cases where both $N_f$ and $N_g$ contribute to the asymmetry generation are possible, they do not bring in new features.

\subsection[Leptogenesis from $N_g$]{\boldmath Leptogenesis from $N_g$}
Here we assume that $N_f$ does not generate any significant amount of $B-L$ asymmetry while it is mainly produced from other RHN $N_g$, which decay promptly as $T < m_{N_g}$. In the limit where $\tilde m_f = m_l$, with $m_l$ being the mass of the lightest active neutrino, the CP violation associated with $N_f$ vanishes. The asymmetry generated from $N_g$ can be parametrized as~\cite{Fong:2013wr}
\begin{equation}
    Y_{B-L} = -\epsilon_g\, \eta_g \, Y_{N_g}^{\rm eq}(0)\,,
\end{equation}
where $Y_{N_g}^{{\rm eq}}(0) = \frac{45}{\pi^{4}\, \gss}$, $\eta_g \leq 1$ is the efficiency factor which takes into account the possible additional effects,%
\footnote{Since a generic seesaw model contains free parameters which cannot be fixed by low energy neutrino observables, we use the efficiency parameter $\eta_g$ to capture possible flavor and washout effects in a model-independent way.}
and $\epsilon_g$ characterizes the CP violation from the decay of $N_g$.
The overall negative sign is a reminder that the $B-L$ asymmetry generated is negative of the lepton asymmetry and to have a positive BAU, one would need to create an excess in the antileptons. Since $N_g$ decays fast, there is a significant washout of the asymmetry generated from inverse decay, and the maximum achievable efficiency is $\eta_g \sim 0.1$~\cite{Giudice:2003jh}. 

In general, the CP parameter is bounded $|\epsilon_g| \leq 1/2$ by the perturbativity condition such that loop corrections do not overwhelm the tree-level amplitude~\cite{Pilaftsis:1997jf}. If $N_{g}=N_{1}$ is the lightest RHN, and under the assumption of hierarchical RHN mass spectrum, the CP parameter is constrained by the Davidson-Ibarra bound~\cite{Davidson:2002qv}
\begin{equation} \label{eq:DI_bound}
    \left|\epsilon_{1}\right| \leq \epsilon_{\rm DI} \equiv \frac{3}{16\pi} \frac{m_{N_1}}{v^{2}} \sqrt{\Delta m_{{\rm atm}}^{2}}\,,
\end{equation}
where we have assumed the lightest active neutrino mass to be much smaller than $\sqrt{\Delta m_{{\rm sol}}^{2}}$ such that $N_{f}$ is not thermalized by neutrino Yukawa interactions. Considering a pair of quasi-degenerate RHNs, the CP parameter can be resonantly enhanced to saturate the perturbativity bound~\cite{Covi:1996fm,Pilaftsis:1997jf}, see Ref.~\cite{Dev:2017wwc} for a recent review on the subject.
Given that $N_g$ should decay fast, we will impose the most optimistic upper bound on the product $|\epsilon_g|\, \eta_g$, $|\epsilon_g|\, \eta_g \lesssim 0.05$, coming from $\eta_g \lesssim 0.1$ and $|\epsilon_g| \leq 0.5$.

After the electroweak sphaleron freezeout, and taking into account of the dilution from the decay of $N_{f}$, we have the final BAU
\begin{equation} \label{eq:B_asymmetry}
    Y_{B}  =\frac{1}{d} \times \frac{30}{97}\, Y_{B-L}
\end{equation}
which has to match the observed value $Y_{B}^{{\rm obs}} \simeq 9 \times 10^{-11}$~\cite{Aghanim:2018eyx}. Since $N_{f}$ is not responsible
for generating the $B-L$ asymmetry, its decay temperature can be as low
as the BBN scale, $T_d > T_\text{BBN}$.
\begin{figure}
	\centering
	\includegraphics[scale=0.51]{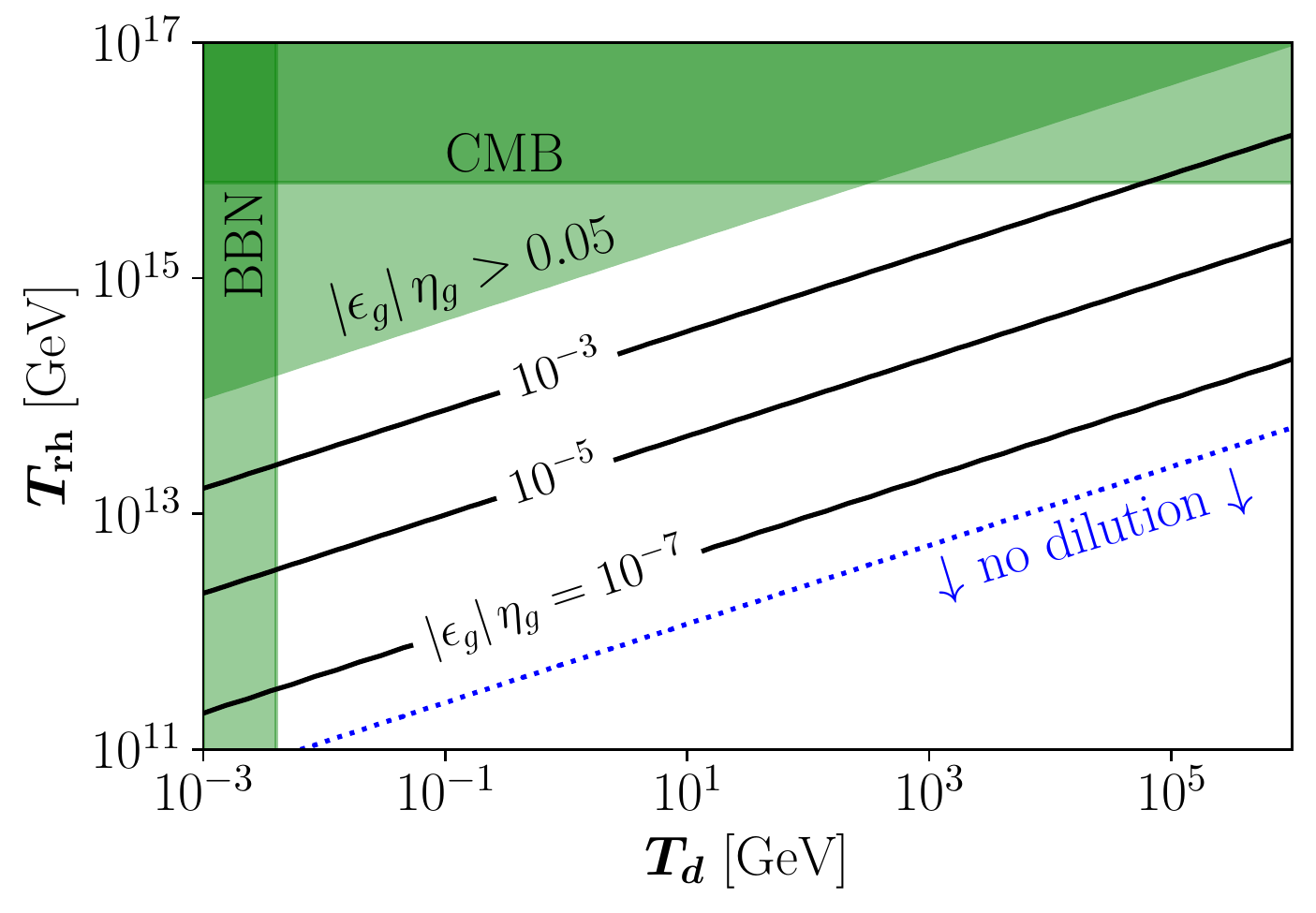}
    \caption{Leptogenesis from $N_{g\neq f}$: Contour lines for the product $|\epsilon_g|\, \eta_g$, showing the parameter space compatible with BAU.
    We have assumed $m_{N_f} = 10^{12}$~GeV, $m_\phi = 3\times 10^{13}$~GeV, and $\Tmax/\Trh = 10^3$.
    Green areas are in tension with CMB or Lyman-$\alpha$ observations, or correspond to $|\epsilon_g|\, \eta_g > 0.05$.
    }
	\label{fig:BAU-Ni}
\end{figure} 
In Fig.~\ref{fig:BAU-Ni}, we show the parameter space of leptogenesis from $N_g$ that generates the observed BAU for various values of $|\epsilon_g|\, \eta_g$, taking $m_{N_f}= 10^{12}$ GeV, $m_\phi = 3\times 10^{13}$~GeV, and $\Tmax/\Trh = 10^3$.
Below the blue dotted line, the dilution factor is $d \simeq 1$, and therefore $N_f$ plays no role in the dilution of the BAU.

\subsection[Leptogenesis from $N_f$]{\boldmath Leptogenesis from $N_f$}
Alternatively, leptogenesis could also come from the decay of $N_f$.
Here we consider the scenario where all the other RHNs $N_{g\neq f}$ do not generate a significant amount of $B-L$ asymmetry and the bulk of the $B-L$ asymmetry is mainly generated through $N_f$, which decays very far from equilibrium. This limiting case can be realized for instance, if $N_{g\neq f}$ are too heavy and cannot be efficiently produced through neutrino Yukawa or gravitational interactions and/or CP violation associated with their decays is suppressed by small Yukawa coupling if they are light enough (can be lighter than $N_f$).
In this case, the asymmetry generated from $N_f$ is given by
\begin{equation}
    Y_{B-L} = -\epsilon_{f}\, Y_{N_{f}}(\Trh).
\end{equation}

There are several differences from the previous scenario. First of all, we impose $T_{d} > 130$~GeV such that leptogenesis occurs before the electroweak sphaleron interactions freezeout~\cite{DOnofrio:2014rug}.
Secondly, the $B-L$ asymmetry generation is completely efficient (i.e., $\eta_f=1$) since the washout due to inverse decays is suppressed when decays occur at $T_{d} \ll m_{N_{f}}$.
Finally, the abundance of $Y_{N_{f}}(\Trh)$ is completely determined by gravitational production from inflatons, Eq.~\eqref{eq:Y0}, as we will see in the following.
As for the CP violation, one cannot have exactly the alignment $\tilde m_f = m_l$ where $m_l$ is the lightest neutrino mass as the associated CP violation vanishes. However, one can still have $\tilde m_f \sim m_l$ while keeping a nonvanishing CP violation.

\begin{figure}
	\centering
	\includegraphics[scale=0.51]{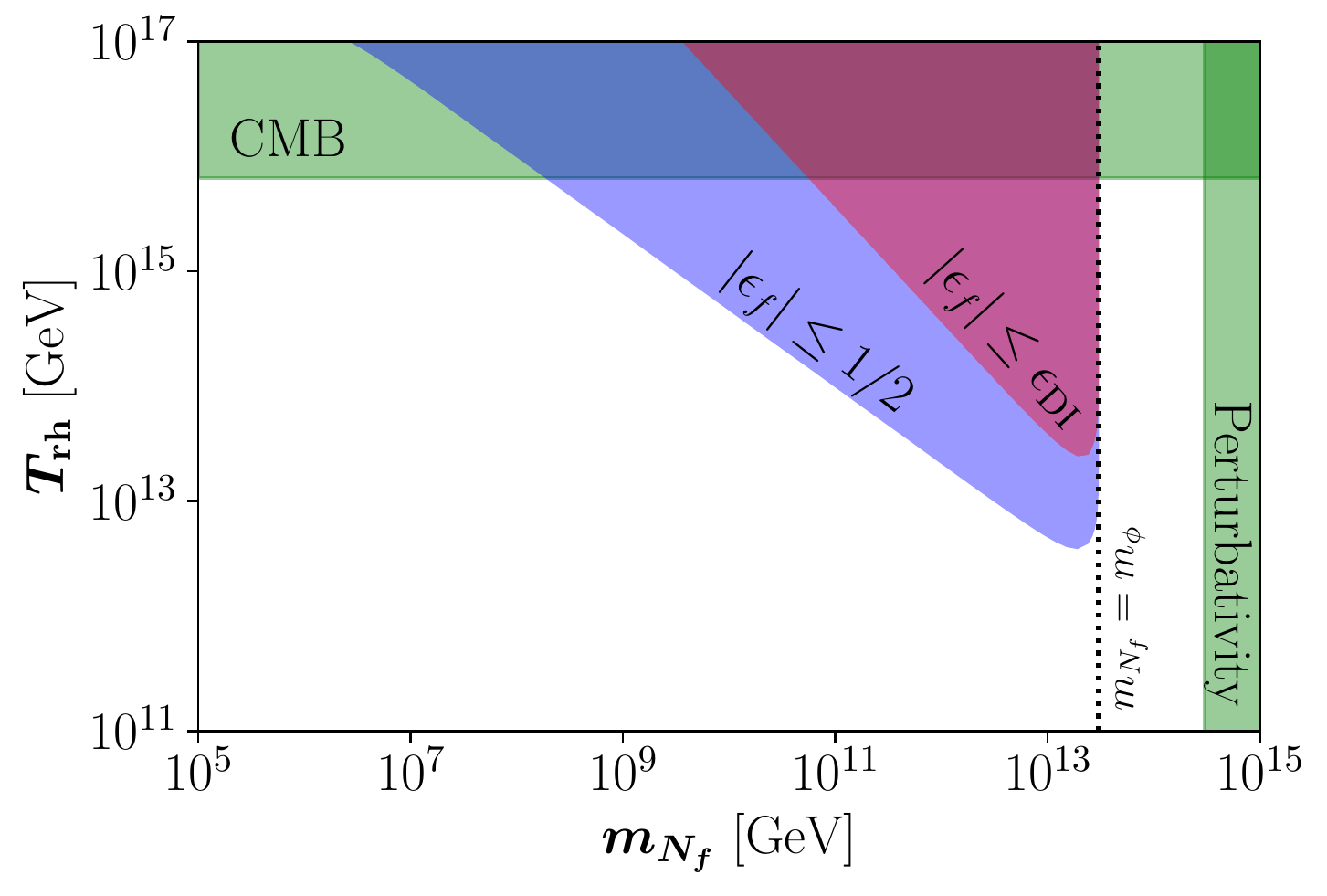}
    \caption{Leptogenesis from $N_f$: Parameter space compatible with BAU, taking $|\epsilon_f| \leq 1/2$ (blue) and $|\epsilon_f| \leq \epsilon_\text{DI}$ (red), $m_\phi = 3\times 10^{13}$~GeV, and $\Tmax/\Trh = 10^3$.
    Green areas are in tension with CMB observations and the perturbativity bound described in the text.}
	\label{fig:BAU}
\end{figure} 
We will consider two representative cases: $\left|\epsilon_{f}\right| \leq 1/2$ and $\left|\epsilon_{f}\right| = \left|\epsilon_{1}\right| \leq \epsilon_{{\rm DI}}$.
For this case of leptogenesis from $N_f$, Fig.~\ref{fig:BAU} shows the parameter space compatible with BAU, taking $|\epsilon_f| \leq 1/2$ (blue) and $|\epsilon_f| \leq \epsilon_\text{DI}$ (red), assuming $m_\phi = 3\times 10^{13}$~GeV and $\Tmax/\Trh = 10^3$.
Even if RHNs can be produced via the two previously mentioned gravitational mechanisms (i.e., inflaton scattering and UV freeze-in), in this framework the observed BAU is dominantly generated with RHNs produced by the annihilation of inflatons.
To produce sufficient $N_f$, we also note that high reheating temperatures are required, $\Trh \gtrsim 10^{12}$~GeV, and RHN with masses $m_{N_f} \gtrsim 10^8$~GeV or $10^{11}$~GeV for $|\epsilon_f| \leq 1/2$ or $|\epsilon_f| \leq \epsilon_\text{DI}$, respectively.

\section{Dark Matter and the Baryon Asymmetry of the Universe} \label{sec:DMBAU}

In this section we study the tight constraints imposed by requiring to match simultaneously the DM relic density and the BAU, in the two scenarios previously described for leptogenesis.

\subsection[Leptogenesis from $N_g$]{\boldmath Leptogenesis from $N_g$}
\begin{figure}
	\centering
	\includegraphics[scale=0.51]{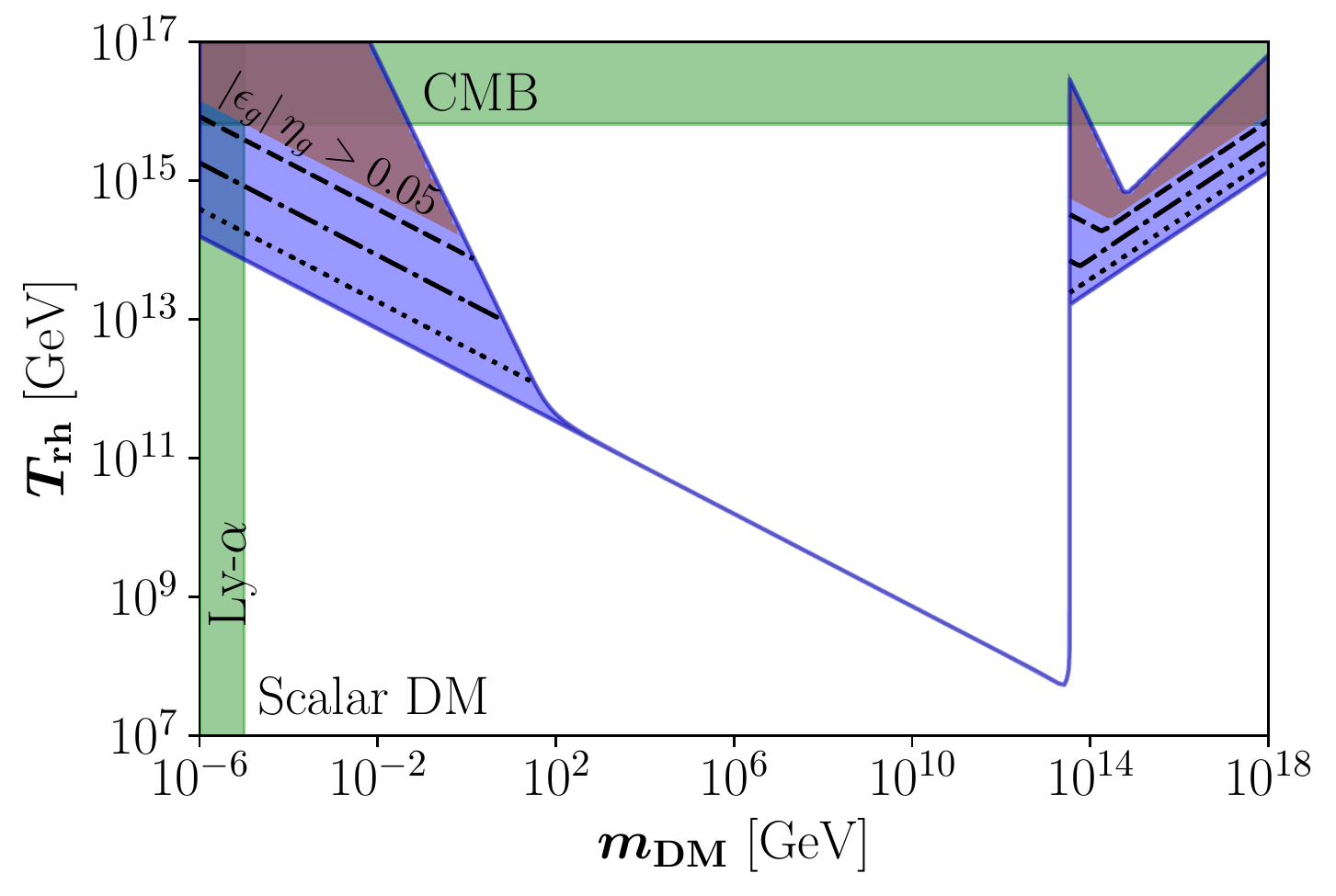}
	\includegraphics[scale=0.51]{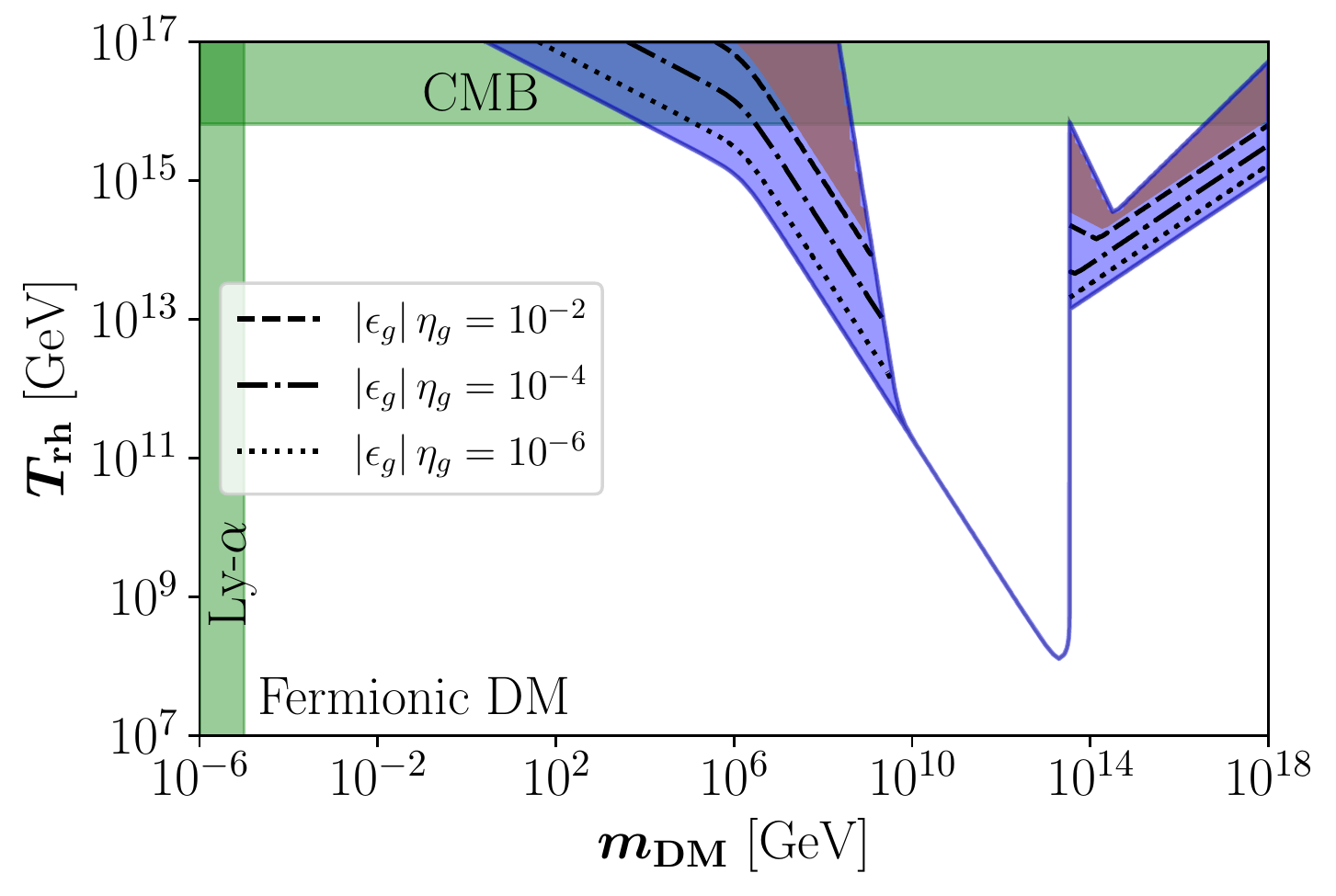}
    \caption{Contour lines for the parameter space reproducing the observed DM relic abundance, for scalar (left panel) and fermionic (right panel) DM, for $\Tmax/\Trh=10^3$, $m_\phi = 3 \times 10^{13}$~GeV, and $m_{N_f} = 10^{12}$~GeV.
    The width of the blue bands show the effect of dilution on the DM abundance.
    Green areas are in tension with CMB and Lyman-$\alpha$ observations.
    }
	\label{fig:DMprod-dil-lepto}
\end{figure} 
In Fig.~\ref{fig:DMprod-dil-lepto}, the colored bands show the parameter space that reproduces the observed DM abundance, for scalar (left panel) and fermionic (right panel) DM,
for $\Tmax/\Trh = 10^3$ and $m_\phi = 3 \times 10^{13}$~GeV and $m_{N_f} = 10^{12}$~GeV, as in Fig.~\ref{fig:DMprod-dil}.
The thickness of the bands corresponds to the effect of dilution due to the decay of the RHN.
Additionally, the figure also shows the contours for the required values of the combination of parameters $|\epsilon_g|\, \eta_g =10^{-2}$, $10^{-4}$ and $10^{-6}$, needed in order to reproduce the observed BAU.
The BAU is over-diluted above these contours.
The dark region corresponding to $|\epsilon_g|\, \eta_g > 0.05$ is beyond the theoretically expected limit, and therefore not considered.

\subsection[Leptogenesis from $N_f$]{\boldmath Leptogenesis from $N_f$}
In this section, the simultaneous production of DM and BAU via gravitational processes is studied.
While DM could be produced by two previously studied mechanisms (i.e., inflaton annihilation and scattering of SM particles, cf. Fig.~\ref{fig:DMprod}), RHNs responsible for the BAU are dominantly produced by gravitational scattering of inflatons, cf. Fig.~\ref{fig:BAU}.

Let us first focus on the case where DM is produced by the scattering of inflatons.
From Eq.~\eqref{eq:Y0}, it is possible to relate the asymmetry generated by $N_f$ to the DM abundance.
One gets that
\begin{equation}
    |Y_B| = \frac{30}{97}\, |\epsilon_f|\, Y_0^\text{DM} \times
    \begin{cases}
        \frac18 \left(\frac{m_{N_f}}{m_\phi}\right)^2 &\text{for scalar DM,}\\[8pt]
        \left(\frac{m_{N_f}}{\mdm}\right)^2 &\text{for fermionic DM,}
    \end{cases}
\end{equation}
independently of the dilution, as it affects both DM and the BAU in the same way.
Using the fact that $Y_B \simeq 9 \times 10^{-11}$ and $Y_0^\text{DM} \, \mdm \simeq 4.3 \times 10^{-10}$~GeV, one has that
\begin{equation} \label{eq:BAUDM1}
    m_{N_f} \simeq
    \begin{cases}
        \frac{2.3}{\sqrt{|\epsilon_f|}} \sqrt{\frac{\mdm}{\text{GeV}}}\, m_\phi &\text{for scalar DM,}\\[8pt]
        \frac{0.8}{\sqrt{|\epsilon_f|}} \sqrt{\frac{\mdm}{\text{GeV}}}\, \mdm &\text{for fermionic DM,}
    \end{cases}
\end{equation}
independently of $\Trh$ and $\Tmax$.

Alternatively, DM could also be produced by gravitational freeze-in via scattering of SM particles.
In the case of DM lighter than $\Trh$, Eq.~\eqref{eq:FIlight} allows to write the asymmetry as
\begin{equation}
    |Y_B| = \frac{\gs^2\, \pi^3}{2979840} \frac{|\epsilon_f|}{\alpha} \left(\frac{m_{N_f}}{m_\phi}\right)^2 \left(\frac{\Tmax}{\Trh}\right)^4 Y_0^\text{DM},
\end{equation}
which implies that
\begin{equation} \label{eq:BAUDM2}
    m_{N_f} \simeq 1.3\, \sqrt{\frac{\alpha}{|\epsilon_f|}}\, \sqrt{\frac{\mdm}{\text{GeV}}} \left(\frac{\Trh}{\Tmax}\right)^2 m_\phi\,.
\end{equation}
In the case opposite case where DM is heavier than $\Trh$, using Eq.~\eqref{eq:FIheavy} the asymmetry can be rewritten as
\begin{equation}
    |Y_B| = \frac{|\epsilon_f|}{198656} \left(\frac{3 \pi^9}{20} \frac{\gs^{12}}{\alpha^3 \gss^4} \right)^\frac17 \left(\frac{m_{N_f}}{m_\phi}\right)^2 \left(\frac{\Tmax}{\Trh}\right)^4 \left(\frac{\mdm}{M_P}\right)^\frac{12}{7} \frac{\left(Y_0^\text{DM}\right)^\frac37}{d^\frac47}
\end{equation}
which implies that
\begin{equation} \label{eq:BAUDM3}
    m_{N_f} \gtrsim \frac{9.4 \times 10^{13} \alpha^\frac{3}{14}}{\sqrt{|\epsilon_f|}} \left(\frac{\text{GeV}}{\mdm}\right)^\frac{9}{14} \left(\frac{\Trh}{\Tmax}\right)^2 m_\phi\,,
\end{equation}
where the inequality comes from the fact that $d \geq 1$.

\begin{figure}
	\centering
	\includegraphics[scale=0.51]{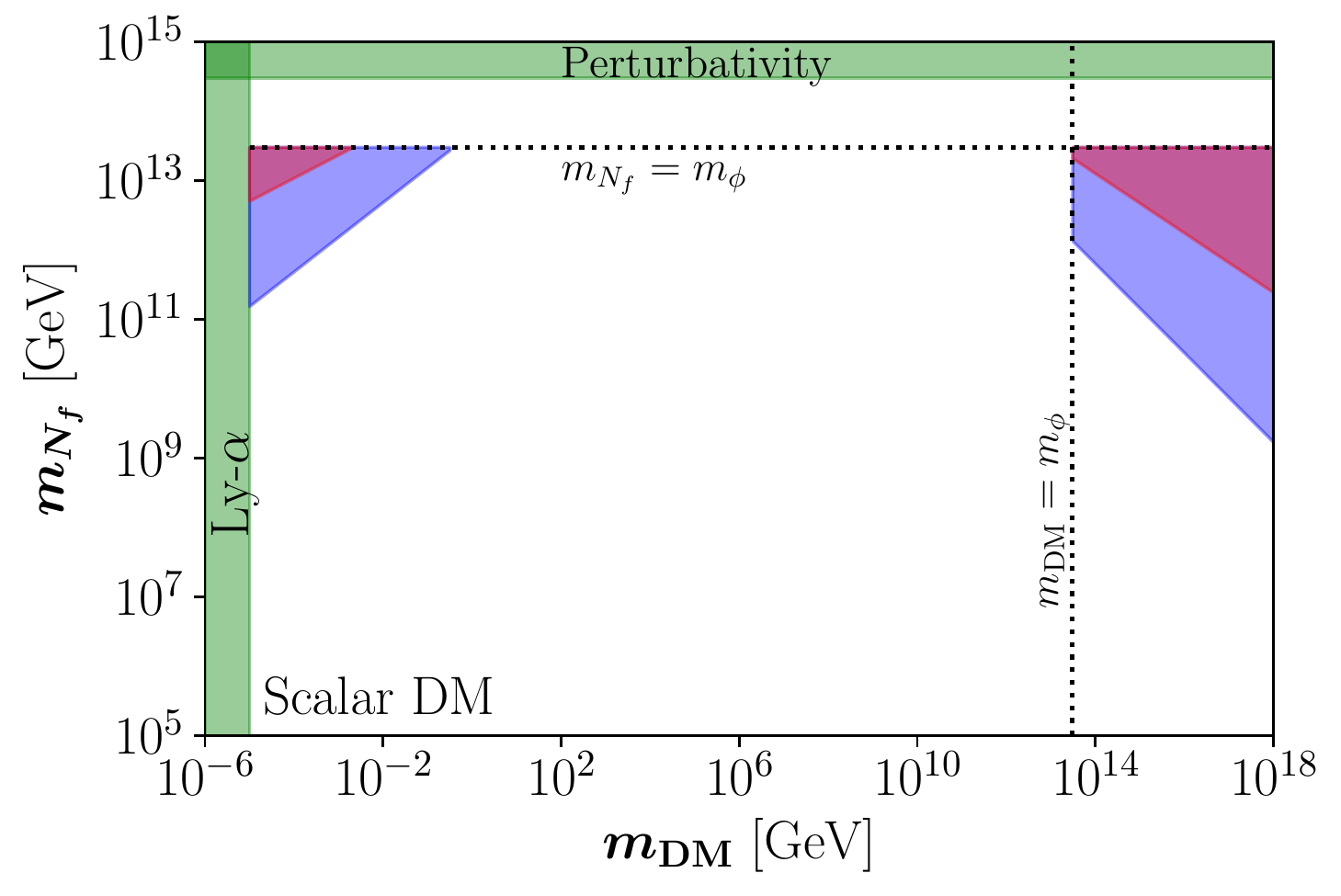}
	\includegraphics[scale=0.51]{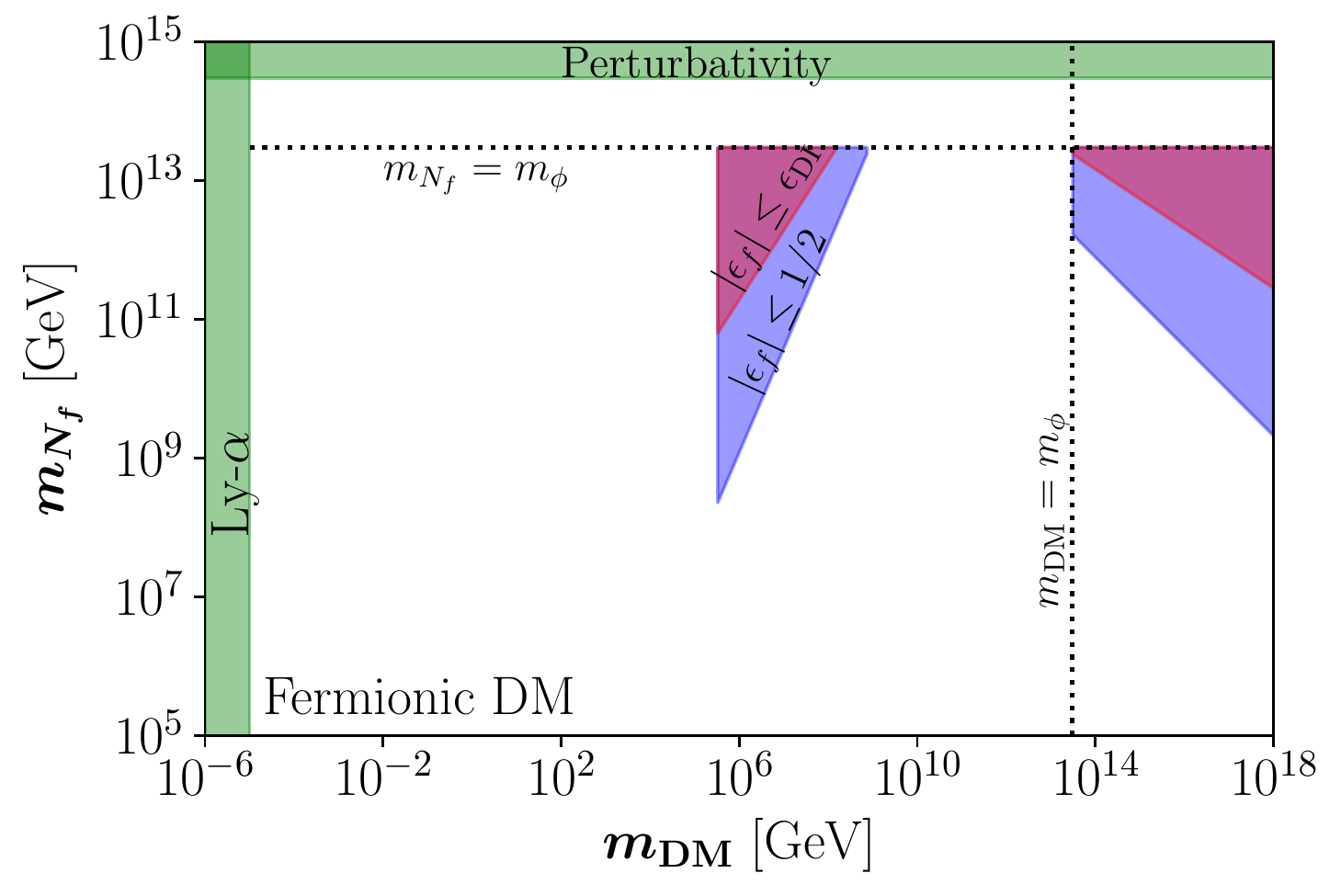}
    \caption{Parameter space compatible with DM and BAU for scalar (left panel) and fermionic (right panel) DM, and $|\epsilon_f| \leq 1/2$ (blue) or $|\epsilon_f| \leq \epsilon_\text{DI}$ (red), for $N_f$ leptogenesis.
    We have assumed $m_\phi = 3\times 10^{13}$~GeV and $\Tmax/\Trh = 10^3$.
    The dotted black lines show $m_{N_f}=m_\phi$ and $\mdm=m_\phi$.
    }
	\label{fig:BAU-DM}
\end{figure} 
Figure~\ref{fig:BAU-DM} shows the parameter space that reproduces simultaneously the BAU and the DM relic abundance, for scalar (left panel) and fermionic (right panel) DM, assuming $m_\phi = 3\times 10^{13}$~GeV and $\Tmax/\Trh = 10^3$.
For leptogenesis from $N_f$, we have assumed two scenarios $|\epsilon_f| \leq 1/2$ (blue areas) and $|\epsilon_f| \leq \epsilon_\text{DI}$ (red areas).
Two separate allowed regions appear, whose existence can be understood as follows.
The first region comes from the DM production out of inflatons ($\mdm < m_\phi$).
The sharp cut at $m_{N_f} = m_\phi$ corresponds to the fact that $N_f$ is produced by inflaton scattering, and therefore the region $m_{N_f} > m_\phi$ is not kinematically allowed.
Additionally, from Fig.~\ref{fig:BAU}, we see that in order to produce enough baryon asymmetry, we need a sufficient $N_f$ abundance which implies $m_{N_f} \gtrsim 10^8$~GeV or $10^{10}$~GeV for $|\epsilon_f| \leq 1/2$ or $|\epsilon_f| \leq \epsilon_\text{DI}$, respectively, and a high reheating temperature $\Trh \gtrsim 10^{12}$~GeV.
This last fact implies that, for inflaton scatterings, one can only accommodate DM with masses $\mdm \lesssim 1$~GeV or $\mdm \lesssim 10^9$~GeV, for scalar or fermionic DM, respectively, cf. Fig.~\ref{fig:DMprod-dil}.
We note that in this case, fermionic DM lighter than $\mdm \lesssim 10^5$~GeV cannot be produced out of the scattering of inflatons.
However, heavier DM with masses $\mdm > m_\phi$, can be generated by gravitational SM scatterings, and therefore a second allowed region appears in the figure.
In both cases, the lower bounds on $m_{N_f}$ come from the requirement of producing simultaneously the observed values of DM and the BAU, as given by Eqs.~\eqref{eq:BAUDM1} and~\eqref{eq:BAUDM3}.

\section{Conclusions} \label{sec:con}

Dark matter (DM) and the baryon asymmetry of the Universe (BAU) are two pressing open problems of the standard model of particle physics (SM).
A common framework for producing the BAU is baryogenesis through leptogenesis, via the decay of right-handed neutrinos (RHNs).
Both the DM and RHNs could have been produced in the early Universe via the standard particle physics portals (Higgs portal, neutrino portal, or kinetic mixing).
However, those states are also unavoidable produced by the irreducible gravitational interaction.
In this scenario, SM gravitons mediate between the dark and visible sectors and produce the whole abundance of DM and RHNs from annihilations of SM particles or inflatons.

In this work, we have considered the gravitational production of DM and a long-lived RHN $N_f$ which couples feebly to the SM sector.
For baryogenesis via leptogenesis two scenarios were studied.
In scenario $(i)$, leptogenesis is driven mainly by RHNs $N_{g\neq f}$ which equilibrate with the SM sector, and the late decay of $N_f$ can dilute both the DM and the generated BAU.
We obtained strong bounds on the product of the CP asymmetry with the efficiency ($|\epsilon_g|\, \eta_g$) required to reproduce the observed DM and BAU as a function of the reheating parameters and the DM mass, cf. Fig.~\ref{fig:DMprod-dil-lepto}.
In scenario $(ii)$, we consider leptogenesis proceeding dominantly from the decay of $N_f$, i.e., $N_f$ produces simultaneously the dilution and the BAU. Since the DM and $N_f$ abundances are both determined from gravitational production, we obtain an nontrivial relations between the DM mass and $m_{N_f}$, cf. Fig.~\ref{fig:BAU-DM}.
As an example, we have chosen the benchmark $m_\phi = 3\times 10^{13}$~GeV and the ratio $\Tmax/\Trh =10^3$.
While $10^8\,{\rm GeV} \lesssim m_{N_f} < m_\phi$, the allowed DM mass falls into two regimes: heavy $\mdm \gtrsim m_\phi$ (independently of spin) from gravitational production out of the scattering of SM particles, and light DM $\mdm \lesssim 1$~GeV (for scalar) and  $10^{5}\,{\rm GeV} \lesssim \mdm \lesssim 10^9$~GeV (for fermion) from production via inflaton scatterings.

All in all, we found that gravitational scattering can simultaneously produce the whole DM abundance and enough RHNs responsible for making baryogenesis via leptogenesis.
This scenario typically requires high reheating temperatures and receive significant boosts for long reheating periods, i.e., large $\Tmax/\Trh$ ratios.
Additionally, while DM can span over a large range of masses (from few keV to the Planck scale), RHNs have to be very heavy, with masses close to the reheating temperature.

Going further in this kind of analysis requires to assume a specific inflationary scenario to fix the inflaton mass, together with $\Trh$ and $\Tmax$.
Additionally, it could give rise to correlated cosmological observables like the spectrum of primordial gravitational waves or of primordial black holes.
Alternatively, one can assume a specific seesaw model in order to have stronger correlations between leptogenesis and neutrino observables.
However, these avenues are beyond the scope of the present study.

\section*{Acknowledgments}
NB received funding from Universidad Antonio Nariño grants 2019101 and 2019248, the Spanish FEDER/MCIU-AEI under grant FPA2017-84543-P, and the Patrimonio Autónomo - Fondo Nacional de Financiamiento para la Ciencia, la Tecnología y la Innovación Francisco José de Caldas (MinCiencias - Colombia) grant 80740-465-2020.
CSF acknowledges the support by FAPESP grant 2019/11197-6 and CNPq grant 301271/2019-4.
This project has received funding/support from the European Union's Horizon 2020 research and innovation programme under the Marie Skłodowska-Curie grant agreement No 860881-HIDDeN.

\bibliographystyle{JHEP}
\bibliography{biblio}

\end{document}